\documentclass[aps,prb,twocolumn,groupedaddress,showpacs]{revtex4-1}
\usepackage{graphicx}
\usepackage{dcolumn}% Align table columns on decimal point
\usepackage{bm}% bold math
\usepackage{color}% colour
\usepackage{url}
\begin{document}

\title{Correlated Electron Pseudopotentials for 3d-Transition Metals}

\author{J. R. Trail}\email{jrt32@cam.ac.uk}
\affiliation{Theory of Condensed Matter Group, Cavendish Laboratory, J
  J Thomson Avenue, Cambridge CB3 0HE, United Kingdom}

\author{R. J. Needs}%\email{rn11@cam.ac.uk}
\affiliation{Theory of Condensed Matter Group, Cavendish Laboratory, J
  J Thomson Avenue, Cambridge CB3 0HE, United Kingdom}

\date{\today}

\begin{abstract}

 A recently published correlated electron pseudopotentials (CEPPs) 
 method has been adapted for application to the 3d-transition metals, 
 and to include relativistic effects.
 New CEPPs are reported for the atoms Sc$-$Fe, constructed from atomic 
 quantum chemical calculations that include an accurate description of 
 correlated electrons.
 Dissociation energies, molecular geometries, and zero-point vibrational 
 energies of small molecules are compared with all electron results, 
 with all quantities evaluated using coupled cluster singles doubles 
 and triples (CCSD(T)) calculations.
 The CEPPs give better results in the correlated-electron calculations 
 than Hartree-Fock-based pseudopotentials available in the literature.

\end{abstract}


% Pseudopotential method (electronic structure of solids), 71.15.Dx
% Monte Carlo methods quantum Monte Carlo, 02.70.Ss
% Electronic structure electron correlation calculations, 31.15.V-

\maketitle

\section{Introduction \label{introduction}}

Pseudopotentials form a useful part of the toolkit of methods for
calculating the properties of interacting atoms in a variety of
environments.\cite{Dolg_2012} They simplify solving the Hamiltonian by
reducing the number of electrons that need to be considered and 
replacing the singular Coulomb interaction between the remaining
electrons and the nuclei with a smooth, although usually non-local,
effective potential.  Here we present pseudopotentials that may be
used with the majority of many-body techniques, although our primary
interest is their use in quantum Monte Carlo (QMC) calculations.
\cite{Ceperley-Alder_1980,Foulkes_QMC_review,Casino_reference,
  Lester_review_2012}

The computational cost of fermion QMC calculations increases as
approximately the third or fourth power of the number of particles,
which is a significant improvement over other correlated wave function
methods.  The scaling with the atomic number $Z$ is, however
$Z^{5}-Z^{6.5}$. \cite{Ceperley_1986,Ma_2005_all_electron_atoms} The
use of pseudopotentials reduces the effective value of $Z$, making QMC
calculations feasible for systems with heavy atoms.

Designing pseudopotentials for 3d-transition metal atoms is of
particular interest.  Obtaining an accurate description of molecules
and bulk systems containing transition metal atoms presents a
important challenge, primarily due to the strong electron-electron
correlation that they exhibit.
Transition metal atoms are particularly
important\cite{organomet,surfsci,catalysis,hitemp,astrophys}
and constructing accurate pseudopotentials for them is demanding.

In a recent paper we presented a new type of correlated electron
pseudopotential (CEPP) \cite{Trail_2013_pseudopotentials} that uses a
many-body generalisation of the preservation of the
independent-electron scattering properties 
that defines the semi-local
norm-conserving pseudopotential.  This generalisation, together with
data from atomic multi-configuration Hartree-Fock (MCHF) calculations,
provides a pseudopotential that includes correlation effects from the
start.\cite{correlatedppots}

The primary limitation of our method is that the pseudopotentials are
generated from systems containing only a single valence electron.  For
almost all elements this restricts us to generating pseudopotentials 
from ions, rather than neutral atoms.
Consequently, we must justify transferring a pseudopotential obtained
from calculations with a single valence electron to systems that are
close to neutral.

Our previous paper \cite{Trail_2013_pseudopotentials} demonstrated
excellent pseudopotential transferability for the atoms H and Li$-$F.
However, the 3d-transition metals place a greater demand on
transferability.  A larger number of electrons must be considered as
valence electrons, resulting in pseudopotentials generated from highly
ionised atoms.  The primary goal of this work is to demonstrate that
CEPPs also exhibit excellent transferability for the 3d-transition
metals.  We will consider our goal to have been achieved if the CEPPs
provide a significant improvement over Hartree-Fock (HF)
pseudopotentials generated from neutral atoms, and reproduce
relativistic all electron (AE) coupled cluster results to within
chemical accuracy.

Pseudopotentials provide an uncontrolled approximation to the
interaction between the core and valence electrons, in the sense that
there is no parameter whose increase causes the error to monotonically
decrease.  The associated uncontrolled error means that it is
necessary to test the performance of a pseudopotential over a wide
range of systems.  We test the CEPPs in reproducing the properties of
a test set of small molecules that contain the 3d-transition metal 
atoms Sc$-$Fe, and H, C, N, O, and F.  (The CEPPs for the first row 
atoms are from Ref.\ \onlinecite{Trail_2013_pseudopotentials}.)

Coupled-cluster calculations with single, double, and perturbative
triple excitations (CCSD(T)), or higher, are required to obtain
quantitatively accurate ground states.  We calculate data at the
CCSD(T) level of theory, with a Gaussian basis, and the
\textsc{MOLPRO}\cite{molpro} code.  Data obtained using the CEPPs are
compared with AE results for the same molecules.  We also compare
results obtained with CEPPs to those obtained using the two well
established libraries of pseudopotentials generated with HF theory,
the norm-conserving Dirac-Fock pseudopotentials of Trail and
Needs\cite{Trail_2005_pseudopotentials,TNDF_website} (TNDF), and the
scalar-relativistic energy-consistent HF pseudopotentials of Burkatzki
\emph{et al.}\cite{Burkatzki_2007,BFD_website} (BFD).

In Sec.\ \ref{CEPPs_theory} we describe the generation of 
the CEPPs.  Subsection\ \ref{theory} summarises the theory used to 
define the CEPPs, while subsection\ \ref{implementation} describes 
the implementation and details of the numerical calculations used to
generate the pseudopotentials.  Section\ \ref{results} presents and
analyses CCSD(T) results for a number of atoms and molecules. Three 
different pseudopotentials are considered, in addition to an AE 
description.  Subsection\ \ref{titanium_atom} examines pseudopotential 
errors for the Ti atom.  Subsection\ \ref{transition_molecules} 
provides a more complete comparison of errors for a range of 
pseudopotentials, molecules, and properties.  The validity of the 
CEPPs in plane wave basis set methods is established for the 
representative case of a TiO$_2$ molecule in Sec.\ \ref{bulkdft_tio2}.
Our conclusions are drawn in Sec.\ \ref{conclusions}.

Atomic units are used, unless otherwise indicated.

\section{Correlated electron pseudopotentials \label{CEPPs_theory}}

The CEPP method may be viewed as a generalisation of standard independent 
electron norm-conserving pseudopotential 
theory\cite{Hamann_pseudopotentials_1979} to the many-body case.
It is based on the preservation of scattering properties for interacting 
electrons, and reduces to the standard norm-conserving pseudopotential 
for non-interacting electrons in an effective one-body potential.

In summary, we start by calculating the many-body wave function for an 
isolated atom, including electron-electron correlation.
We represent this $p$-electron fully correlated AE atom 
as a $p$-body density matrix.
A spherical `core-region' is then defined, and the $p$-body density 
matrix is reduced to an $n_v$-body density matrix outside of this core 
region, where $n_v$ is the number of valence electrons.
The density matrix within the core region takes a model form, chosen
such that the entire $n_v$-body density matrix represents the ground
state of an $n_v$-electron pseudo-atom.
The CEPP is then constructed from this pseudo-atom by inversion of the
density matrix to obtain the effective potential whose ground state
solution is the pseudo-atom itself.  This approach was previously
found to provide accurate pseudopotentials for first row
atoms. \cite{Trail_2013_pseudopotentials}
Relativistic effects are expected to be more important for the 
3d-transition metal atoms, and therefore we have included them in the 
AE atom, and in the core represented by the CEPP.

\subsection{Theoretical basis \label{theory}}

The density matrix of the AE atom with $p$ electrons is denoted
$\Gamma^p(\mathbf{r}_1\ldots\mathbf{r}_p ;
\mathbf{r}_1'\ldots\mathbf{r}_p')$, and a density matrix that
represents the same scattering properties to order $n_v$ is given by
the reduced density matrix,\cite{Trail_2013_pseudopotentials}
\begin{widetext}
\begin{equation}
\Gamma^{n_v}(\mathbf{r}_1\ldots\mathbf{r}_{n_v} ; \mathbf{r}_1'\ldots\mathbf{r}_{n_v}') =
  {p \choose n_v} \int d\mathbf{r}_{n_v+1} \ldots d\mathbf{r}_p
  \Gamma^p( \mathbf{r}_1 \ldots\mathbf{r}_{n_v} ,\mathbf{r}_{n_v+1}\ldots\mathbf{r}_p ;
            \mathbf{r}_1'\ldots\mathbf{r}_{n_v}',\mathbf{r}_{n_v+1}\ldots\mathbf{r}_p )
\end{equation}
\end{widetext}
where the co-ordinates with indices $1$ to $n_v$ are outside of the
core region of the isolated atom.
A generalisation of the one-body L\"uders relation\cite{Luders_1955}
to the many-body case shows that having the correct $n_v$-body density
matrix outside of the core region is sufficient to preserve the
scattering properties.
Details are available in the literature.\cite{Acioli_1994}

The core region is defined as a sphere of radius $r_c$, centred on the
nucleus.
Here we limit ourselves to a single valence electron, so the required
condition corresponds to conserving the charge density outside of the
core region.

The density of the $1$-electron pseudo-atom is defined as 
\begin{equation}
\label{den2}
\rho(\mathbf{r}) = \left\{ \begin{array}{ll}
  \sum_{i>n_c} o_i\psi^2_i(\mathbf{r}) &  |\mathbf{r}| \geq r_c \nonumber \\
  \phi^2(\mathbf{r})                   &  |\mathbf{r}| < r_c ,
\end{array} \right.
\end{equation}
with $\{\psi_i\}$ the natural orbitals (NOs) of the AE atom ordered such that 
the associated occupation numbers, $o_i$, are monotonically decreasing with $i$.
The pseudo-density in the core region, $\phi^2$, is given by the standard 
Troullier-Martins form.\cite{Troullier_1991}
The first $n_c$ NOs are excluded to ensure that the CEPP is equivalent to the 
norm-conserving pseudopotential for non-interacting electrons, and 
excluding these `core' NOs provides a pseudo-atom with the same scattering 
properties as the AE atom modified such that the core is entirely 
contained within the core region.
The pseudopotential is then obtained as the effective potential that 
results from inversion of the radial Schr\"odinger equation (SE), with 
the same spatial angular momentum quantum number as the AE state.

We divide space into three regions.
Within the asymptotic region $III$, defined by $r_0 \leq r$, the potential 
due to the excluded core is described by a charge of $Z-2n_c$, and a 
dipole polarizability $\alpha$.
Region $II$ is defined by $r_c \leq r < r_0$, and contains the potential 
obtained by direct inversion of the SE.
Continuity of the potential at $r=r_0$ provides the eigenvalue used in 
the inversion.
Within the core region $I$, defined by $0 \leq r < r_c$, the potential is 
again obtained by direct inversion of the SE, using the same eigenvalue as 
in region $II$.
The ground state solution of a one electron SE with the resulting
effective potential, $V(r)$, then reproduces the charge density of the
pseudo-atom accurately, with a negligible error due to the asymptotic
form used in region $III$.

Since we are considering relativistic states, a different potential
channel arises for distinct $J$ quantum numbers of the underlying AE
state.  Each set of channels with the same spatial angular momentum 
quantum number, $L$, is then weighted by $(2J+1)$ and averaged, to 
provide the averaged relativistic effective potential (AREP) for use 
in a non-relativistic Hamiltonian.\cite{Kleinman_1980,Bachelet_1982} 
(Spin-orbit coupling can be taken into account by employing the 
appropriate differences, but this has not been addressed here.)
Furthermore, we subtract the one-body part of the semi-empirical core
polarisation potential (CPP),\cite{Shirley_1993,Muller_1984} so that
the one-body part of the core-valence interaction potential is
entirely \emph{ab initio} when the pseudopotential is applied together
with the CPP.

Denoting the potential arising from ground states with spatial angular 
momentum $L$ as $V_{J=L \pm 1/2}$, the $V_l$ channel of the CEPP is 
then given by
\begin{equation}
V_l(r) = \frac{ l V_{J=l-1/2} + (l+1) V_{J=l+1/2} }{ 2l+1 } + \frac{1}{2}\frac{\alpha}{r^4} f(r/\bar{r}_l)
\end{equation}
where $f(x)=(1-e^{-x^2})^2$ is the short range truncation function
used to remove the nonphysical singularity at $r=0$, $\bar r_l$ are
empirically defined cutoff radii, and $\alpha$ is the same polarizability
as used in region $III$.

These channels are then applied as the semi-local operator
\begin{equation}
\label{semiloc}
\hat{V}= \sum_0^{l_{max}-1} \sum_{m=-l}^{m=l}
| Y_{lm} \rangle \left[ V_l(r) - V_{l_{max}}(r) \right] \langle Y_{lm} | + V_{l_{max}}(r) .
\end{equation}

\subsection{Implementation \label{implementation}}

Application of the method described above to generate CEPPs requires
the NOs for the isolated ion.  Gaussian basis calculations for
isolated atoms are unsuitable, since the exponential asymptotic
behaviour of the NOs is vital to ensure that the inversion process
provides the correct asymptotic limit.  We use the \textsc{ATSP2K}
MCHF code\cite{atsp2k} that describes the correlated wave-function as
a multideterminant expansion with orbitals that are tabulated on
a radial grid and free to relax.  In order to describe core-valence 
and intra-core correlation, core electron excitations are
included in the active space (AS) for these calculations.
Relativistic effects are included by performing a post-MCHF
Configuration-Interaction (CI) calculation using a Hamiltonian that
includes Breit-Pauli terms.\cite{Fischer_1997}
The \textsc{ATSP2K} package also provides the AE NOs tabulated on a 
numerical grid, together with the associated occupation numbers.

Only a finite number of orbitals can be included in the calculation, 
which are indexed by an angular momentum eigenvalue $l$, and an 
index analogous to the primary quantum number of hydrogenic atoms, $n$.
For all atomic calculations we have employed the largest ranges of $n$
and $l$ possible without linear dependency problems preventing
convergence.  In practice, this provides $n \le 7$ and $l \le 6$, and
the removal of a $7s$ and $7p$ orbital.  These ranges, together with
single and double excitations of both the core and valence electrons,
define the AS used for the 3d-transition metal ions considered.

In almost all applications we will seek to transfer the pseudopotentials 
to systems that are far closer to neutral than the generating ionic 
states.  Consequentially, we modify the core in the MCHF calculations 
to represent that of a neutral atom, rather than the generating ion.
Freezing the core of the ion to that of the neutral atom is achieved 
via two AE MCHF atomic calculations.
First `relaxed core' MCHF results are generated for the neutral atom, 
with all orbitals and determinant expansion coefficients free to vary.
From the resulting orbitals we select the $n_c$ core orbitals.
Next, we perform a `fixed core' MCHF calculation by including the 
neutral atom core orbitals in the active space (AS), not allowing 
them to relax.
In this second calculation, all determinant expansion coefficients are 
free to vary, but only those orbitals that are not `fixed core' are 
free to relax.
This second MCHF calculation provides the final set of NOs, 
$\{\psi_i \}$.
Note that the second calculation provides an ion with the same 
core orbitals as the neutral atom, but with different core NOs.
When the NOs are used in Eq.\ (\ref{den2}), they provide a pseudo-atom and 
pseudopotential with the core fixed to that of the neutral atom.
A complete AS is equivalent to allowing core orbitals to relax, 
therefore our fixed core orbitals can be viewed as 
defining a finite AS biased towards the neutral atom.

The most important physical parameters are those that define 
the core itself, that is the number of core electrons, $2n_c$, 
the core radius, $r_c$, and the CPP parameters, $\alpha$ and $\bar{r}_l$.
The inadequacy of a frozen [Ar] core ($2n_c=18$) is well established 
for AE calculations and for defining pseudopotentials for the 
3d-transition metals,\cite{Pacios_1988} hence a [Ne] core ($2n_c=10$) 
is used here.

The core radius, $r_c$, should be large enough that the core density
left out of Eq.\ (\ref{den2}) is negligible outside of it.  We take
$r_c$ to be the radius containing $75\%$ of the core charge excluded
from Eq.\ (\ref{den2}), multiplied by $1.9$ for non-local channels,
and by $2.1$ for the local channel.  This empirical rule was obtained
by noting that the transferability of the pseudopotential is expected
to worsen with increasing $r_c$.  With this in mind, we found the
largest core radii (and shallowest pseudopotential) for Ti such that
the performance of the CEPP in TiO$_2$ has not yet begun to
deteriorate, and scaled this with the core radius for other atoms.
For all atoms considered fewer than $7 \times 10^{-4}$ core 
electrons fall outside of the core region.

Parameterised CPPs are required for both the generation and application of 
the CEPPs.
We employ the Shirley-Martin\cite{Shirley_1993} parameterisation, although
these authors do not supply core-polarizabilities and associated $\bar{r}_l$ 
parameters for the [Ne] cores of the 3d-transition metal atoms.
The inversion process used in generating the CEPPs can provide the core 
polarizability from the asymptotic form of the CEPP.
However, parameterising the asymptotic behaviour is prone to numerical error, 
so instead we use the core polarizabilities calculated semi-classically by 
Patil\cite{Patil_1985} that agree well with experimental values.
These polarizabilities also agree well with the asymptotic form of the CEPPs.

We define $\bar{r}_l$ to be the same for each $l$, and equal to the
radius containing $75\%$ of the core charge excluded from Eq.\
(\ref{den2}).  The particular value chosen can be expected to have a
minimal effect on the performance of the CEPP+CPP, since the one-body
part of this potential does not depend on $\bar{r}_l$.

The remaining parameters are defined by requiring that the associated 
errors are small.
The local channel, $l_{max}$, in Eq.\ (\ref{semiloc}) is defined such 
that the orbitals present in the dominant configuration of the neutral 
atom ground state have a maximum spatial angular momentum quantum 
number of $l_{max}-1$.
For the 3d-transition metal atoms this criterion provides CEPPs with 
s, p, d, and f channels, and a local f channel.
The s, p, d, and f channels are generated from the lowest energy 
AE states with terms
$^2S_{\frac{1}{2}}$, $^2P_{\frac{1}{2}}$, $^2P_{\frac{3}{2}}$, 
$^2D_{\frac{3}{2}}$, $^2D_{\frac{5}{2}}$, $^2F_{\frac{5}{2}}$, 
and $^2F_{\frac{7}{2}}$. 

For the transition metal atoms with small cores, $r_0$ plays a more
important role than in the first-row. This occurs due to the 
inclusion of fixed core orbitals from the neutral atom in the AS of 
the ion, and because the asymptotic decay of orbitals in a many-body
atomic system is given by\cite{Katriel_1980}
\begin{equation}
\psi \asymp r^\beta e^{-\sqrt{2I}r},
\end{equation}
where $I$ is the first ionisation energy, and $\beta$ is a constant
that depends on the NOs.

Since neutral core orbitals are present in the ion AS, our AE
calculation provides NOs that exhibit a mixture of the exponential
decay associated with the neutral atom and the ion.
The ionisation energy of the neutral atom is
usually less than that of the ion, and hence the exponential decay of
the neutral atom eventually dominates.  In terms of the potential
arising from inversion of the Schr\"odinger equation, this manifests
itself as a large non-physical smooth step at large radii,
corresponding to a pseudopotential with an incorrect asymptotic form.
(For core orbitals that are free to relax, such a non-physical step
does not occur and the asymptotic behaviour is correct.)

\begin{figure}[t]
\includegraphics[scale=1.00]{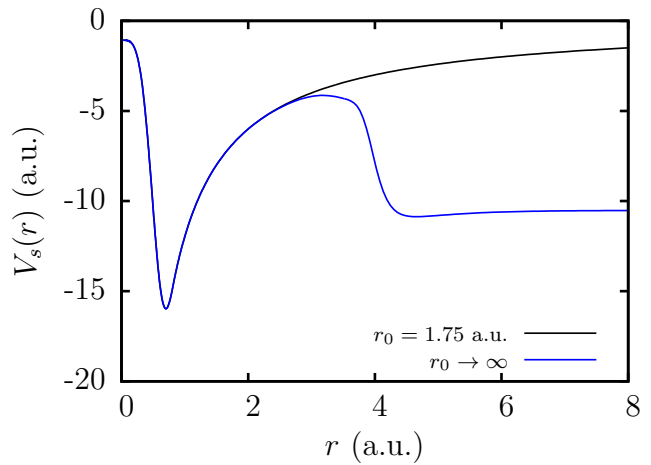}
\caption{ \label{fig1}
The s channel of a CEPP pseudopotentials generated from Ti$^{+11}$ 
for both finite $r_0=1.75$ a.u., and without a region III ($r_0 \rightarrow \infty$).
For $r_0 \rightarrow \infty$, the eigenvalue used in the inversion is not 
available, so we employ the finite $r_0$ value.
On the scale of this figure a CEPP constructed without fixed core orbitals is 
indistinguishable from the finite $r_0$ result.
}
\end{figure}

Figure~\ref{fig1} shows an example of this behaviour for the Ti$^{+11}$ CEPP 
generated with $r_0 \rightarrow \infty$, for which region $III$ is absent.
Just before this smooth step occurs the CEPP shows the expected, and physically 
realistic, asymptotic behaviour, hence the most straightforward way to remove 
this anomalous step is to choose an appropriate value of $r_0$.
Setting $r_0$ as the radius at which the deviation of the CEPP from the	
asymptotic potential is smallest provides an effective solution, 
resulting in values of $r_0$ varying between $0.8 - 2.4$ a.u.\ over all 
channels and atoms considered.
For all atoms considered fewer than $2 \times 10^{-4}$ core 
electrons fall in the region $r>r_0$.
Figure~\ref{fig1} also shows the corrected Ti$^{+11}$ CEPP generated with such 
a choice for the inner radius of region $III$.

Note that for the first row atoms such a non-physical smooth step was 
also present, but was considerably more shallow, and occurred much further from 
the core than for the very ionised transition metal atoms.
This weaker step was indistinguishable from the unavoidable numerical noise 
in the MCHF orbitals far from the nucleus, so there was no need to 
consider it as separate from this numerical error.

To allow these pseudopotentials to be used in a variety of applications, we 
require them to be expressed as an expansion in Gaussian functions.
We employ the same Gaussian expansion and optimisation procedure 
used previously.\cite{Trail_2013_pseudopotentials}

Before commencing with CEPP generation and testing, we note some of its 
properties.
Although the CEPP is defined in terms of the charge density, this definition 
does not invoke KS-DFT (Kohn-Sham Density Functional Theory), hence there is 
no error due to the non-linearity of exchange correlation functionals.
This is an important advantage of the CEPP approach over traditional 
KS-DFT pseudopotentials, since a large part of the transferability error 
for the latter appears to be due to this non-linearity.\cite{Porezag_1999,Zhu_2013}

For mean-field pseudopotentials the variation in the effective
potential due to self-consistency introduces an error in scattering
properties that is neglected by norm-conservation, even at $1^{st}$
order.\cite{Trail_2005_asymptotic}
An advantage of explicitly including correlation effects in the definition 
of the pseudopotential is that such an error does not occur, since the 
Hamiltonian is not defined self-consistently.

The core that is removed in our definition of the CEPP is not unique.
This is unavoidable for a many-body description of an atom.  However,
defining the core in terms of the NOs with the highest occupation
number is a natural generalisation of the core definition used in the
familiar norm-conserving pseudopotential method.
Self-consistent orbitals and NOs are equivalent for independent 
electrons, both provide the density in a diagonal form 
(see Eq.\ (\ref{den2})), and the NOs provide the most rapidly 
convergent multideterminant expansion of all choices of 
orbitals.\cite{Davidson_1972}

Perhaps the least predictable error in the CEPPs arises from the transfer 
of the pseudopotential from a highly ionic atom to a neutral system.
From the generation procedure, we can expect this transferability error to 
be smaller than in KS-DFT, since our CEPPs reproduce many-body 
scattering properties and are not vulnerable to the same exchange-correlation 
errors as KS-DFT pseudopotentials.
Our approach of generating the pseudopotential by introducing the core of a 
neutral atom can be expected to further improve this transferability.

Finally, it is worth noting that our definition of the CEPP in terms of an 
ion with a single valence electron is not a requirement of the underlying 
theory: in general the CEPP may be defined as the potential that results 
in the $n_v$-body pseudo-density matrix for an atom with $n_v$ valence 
electrons.
We limit ourselves to a single valence electron to ensure that the 
required inversion process is computationally achievable, and that 
the resulting pseudopotential takes the form of a one-body semi-local 
pseudopotential.
For the more general case, a $n_v$-body fully non-local potential is 
expected to underlie the $n_v$-body pseudo-density matrix.

\section{Results \label{results}}

\begin{table*}[t]
\begin{tabular}{llcccccccccccrr}                                                                \\ \hline \hline
Atom & $Z_v$ & & \multicolumn{4}{c}{ $r_c$(a.u.) } & & \multicolumn{4}{c}{ $r_0$(a.u.) } & & $\alpha$(a.u.) & $\bar{r}_l$(a.u.)    \\
     &       & & $l=s$ & $p$  & $d$  & $f$  & & $l=s$ & $p$  & $d$  & $f$  & &        &        \\ \hline
Sc   &  11   & & 0.94  & 0.96 & 0.94 & 1.05 & & 1.83  & 1.06 & 1.83 & 1.51 & & 0.0136 & 0.52   \\
Ti   &  12   & & 0.90  & 0.90 & 0.90 & 1.00 & & 1.75  & 1.00 & 1.75 & 2.40 & & 0.0106 & 0.50   \\
V    &  13   & & 0.86  & 0.86 & 0.86 & 0.96 & & 1.67  & 0.90 & 1.42 & 2.30 & & 0.0084 & 0.48   \\
Cr   &  14   & & 0.81  & 0.81 & 0.81 & 0.90 & & 0.96  & 0.90 & 1.35 & 2.17 & & 0.0067 & 0.45   \\
Mn   &  15   & & 0.78  & 0.78 & 0.78 & 0.87 & & 1.52  & 0.80 & 1.23 & 1.20 & & 0.0055 & 0.43   \\
Fe   &  16   & & 0.76  & 0.73 & 0.76 & 0.84 & & 1.47  & 0.84 & 1.18 & 2.02 & & 0.0045 & 0.40   \\ \hline \hline
\end{tabular} \  \
\caption{ \label{tab:1}
Parameters used to define the CEPPs for Sc$-$Fe.
The core radii, $r_c$, and $Z_v=Z-2n_c$ define the core.
The core polarizability, $\alpha$, and associated cutoff radii, $\bar{r}_l$, 
do not change the form of the one-body part of the sum of CEPP and CPPs.
The core polarizabilities are those provided by Patil,\cite{Patil_1985} with 
the $\bar{r}_l$ values obtained as described in the text and taken as equal for 
all $l$.
}
\end{table*}

CEPPs generated as described in Sec.\ \ref{CEPPs_theory} provide the 
four channel parameterised pseudopotentials for the transition metal 
atoms Sc$-$Fe.
Each CEPP is accompanied by a parameterised CPP.
For each atom, the parameters that specify the core properties in the 
construction of the CEPPs, and the CPPs, are given in 
Table~\ref{tab:1}.
We test these CEPPs, together with previously constructed first-row atom 
CEPPs,\cite{Trail_2013_pseudopotentials} using CCSD(T) calculations.
Large Gaussian basis sets are used for both the pseudopotential and AE 
systems.

Experimental data are lacking for many transition metal molecules.
When such data are available, estimated experimental errors are often
absent, or different values for the same quantity differ by significantly
more than the estimated errors, and by more than chemical accuracy.

Furthermore, it is possible for a single-reference CCSD(T) description
to miss a significant fraction of the correlation energy.
Pseudopotentials are designed to provide an effective Hamiltonian that
reproduces the properties of the AE Hamiltonian, hence we can expect
that the same fraction of valence correlation will be missing from a
CCSD(T) calculation with either AEs or CEPPs. 
We therefore concentrate on testing the accuracy of the CEPPs in terms 
of their agreement with AE CCSD(T) results, rather than with 
experimental results.

Relativistic effects are included in the AE CCSD(T) calculations by employing 
the second order Douglas-Kroll-Hess Hamiltonian.
For all atoms and molecules considered we used uncontracted Gaussian basis sets, 
employing aug-cc-pVnZ basis sets\cite{basis} for the pseudopotential calculations, 
and the very similar aug-cc-pVnZ-DK basis sets\cite{basisDK} for the relativistic 
AE calculations.
This choice of basis sets ensures that the AS for the AE and 
pseudopotential calculations are consistent, and not biased towards either. 
(Different contracted basis sets are available for AE, 
TNDF,\cite{TNDF_website} and BFD\cite{BFD_website} calculations.)

The geometry of each molecule was obtained by minimisation of the
total energy using the $n=Q$ basis set.

Dissociation energies were evaluated from total energies for 
atoms and molecules in the optimum geometry, using both 
the $n=Q,5$ basis sets.
Extrapolation of the total energies to the complete basis set limit was then 
performed as in Ref.\ \onlinecite{Trail_2013_pseudopotentials} to 
provide dissociation energies, $D_e$, and estimates of the error from
extrapolation.

A minor difficulty arises due to the implementation of the CPP in 
\textsc{MOLPRO} being limited to smaller basis sets only, those for 
which $l\leq4$ for all basis functions.
This limits the basis sets usable in our CPP calculations to $n \leq T$.
For geometry optimisation we simply choose to exclude the CPP 
potential, since the basis set error for $n=T$ was found to be larger 
than the error due to neglect of the CPP.
For the total energies we extrapolate to the full basis set limit with 
$n=Q,5$ and no CPP, and correct for the absence of the CPP using results 
from $n=T$.
This provides the dissociation energy in the complete basis set limit, 
corrected for the CPP, as  
\begin{equation}
D_e[CPP] = D_{e,est} + D_{e,n=T}[CPP] - D_{e,n=T},
\end{equation}
where $D_{e,est}$ is the result from extrapolating data calculated 
without a CPP to the complete basis set limit.
The dissociation energies, $D_{e,n=T}[CPP]$ and $D_{e,n=T}$, 
include and exclude the CPP, respectively, and result from calculations 
using the $n=T$ basis sets without extrapolation.
Although this limitation on the available basis sets only arises when the 
CEPP+CPP combination is employed, we use the same basis sets for all 
calculations to ensure that errors are as consistent as possible.

In the rest of this section we present an analysis of the performance 
of the CEPPs, quantified by the agreement with a baseline of relativistic 
AE results.
Results from the relativistic HF pseudopotentials available in 
the literature, specifically the norm-conserving TNDF pseudopotentials 
and the energy-consistent BFD pseudopotentials, are also compared.
Note that for the 3d-transition metals the TNDF pseudopotentials have
a larger [Mg] core, whereas the BFD pseudopotentials have the same
[Ne] core used for the CEPPs.

\subsection{Titanium ionisation and excitation energies \label{titanium_atom}}

\begin{figure}[t]
\includegraphics[scale=1.00]{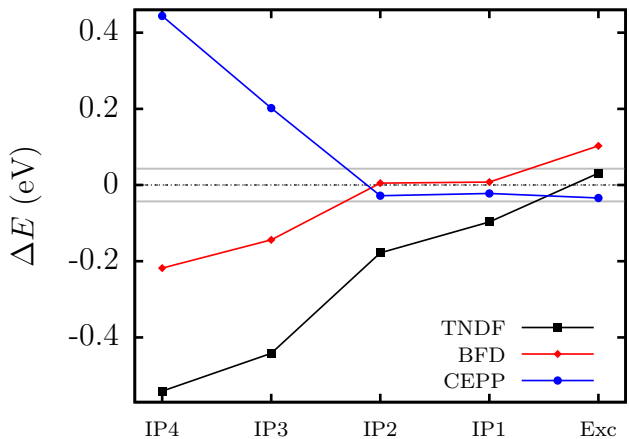}
\caption{ \label{fig2}
Difference between atomic Ti ionisation and excitation energies from 
pseudopotentials, and those from relativistic AE calculations. 
Both are evaluated using CCSD(T), uncontracted aug-cc-pVnZ(-DK) basis sets, 
including excitations of all electrons, and extrapolation to the complete basis 
set limit using $n=Q,5$.
IP4 to IP1 are the energy differences between the 
$^1$S\ [Ar] ,
$^2$D\ [Ar]d$^1$ ,
$^3$F\ [Ar]d$^2$ ,
$^4$F\ [Ar]s$^1$d$^2$ , and
$^3$F\ [Ar]s$^2$d$^2$ states.
Exc is the energy differences between the 
$^3$F\ [Ar]s$^2$d$^2$ and $^5$F\ [Ar]s$^1$d$^3$ states.
The horizontal grey lines indicate the boundaries for chemical accuracy.
}
\end{figure}

We begin with the isolated Ti atom, in order to demonstrate the error 
arising from generating CEPPs from highly ionised atoms and applying them 
to isolated atoms that are close to neutral.
These results provide a useful check on the accuracy with which CEPPs
reproduce AE atomic properties when transferred over the wide energy
range between the generating states and the physically relevant more
nearly neutral states.
We consider the first four ionisation energies and first excitation energy 
of the Ti atom, obtained using CCSD(T).
The deviation of the results for the three pseudopotential types from the
relativistic AE excitation energies are shown in Fig.~\ref{fig2}.
Errors in energy differences are estimated from extrapolation as less than 
$0.02$ eV for all data.

The CEPP clearly demonstrates an accurate reproduction of the AE 
energies for the first two ionisation energies, and the first excitation 
energy, with all three agreeing with relativistic AE values to within 
chemical accuracy.
The agreement is noticeably worse for the other ionisation energies.

There is a systematic error due to the CEPP representation of the core as 
that of a neutral atom. 
While both the fixed core and the finite basis set introduce errors that 
are expected to increase monotonically with increasing ionisation, 
both HF and MCHF AE calculations with core orbitals fixed to those of
Ti ($^2$D) provide ionisation energies that deviate from the equivalent 
relaxed core values by less than chemical accuracy.
From this we conclude that, for the most ionised states, the transferability 
error is dominant.
As we would expect, this error becomes more apparent as the energy 
differences that define the ionisation energies becomes larger.
However, for the states close to neutral this error is less than chemical 
accuracy.

Of the three pseudopotentials considered, the TNDF pseudopotential
shows the largest error, most likely due to the inclusion of the 3s
electrons in the core.
The BFD pseudopotential is designed to reproduce all of the energy differences 
shown in the figure, at the HF level of theory.
As we might expect, for IP1 and IP2 the BFD pseudopotential reproduces 
AE CCSD(T) values to within chemical accuracy.
However, it fails to achieve this for the other energy differences shown.
Unlike either the TNDF or BFD pseudopotentials, the CEPP provides
better than chemical accuracy for the first two ionisation energies
and the first excitation energy.  This suggests that the CEPP may be
the best choice, even though the BFD pseudopotential provides more
accurate values for IP1 and IP2.

It seems reasonable to conclude that for isolated atoms the CEPP and BFD 
pseudopotentials show a similar level of accuracy, that the CEPP is 
marginally more reliable, and that the CEPP transfers well from the highly
ionised generating state to close-to-neutral atomic states, providing 
better than chemical accuracy for energy differences between atomic states 
close to neutral.

It is possible to construct further tests by 
comparing the components of the CCSD(T) total energies for  
atoms, excluding core and core-valence contributions for the  
AE case.  However, such a division of the AE total energy 
into core, core-valence, and valence parts will generally
not be consistent with the NO definition of the core used 
to define the CEPPs, and such tests are not used here.

\subsection{Small 3d-Transition metal molecules \label{transition_molecules}}

\begin{table*}[t]
\begin{tabular}{ p{3em}p{3em} p{3em}p{3em} p{3em}p{3em} p{3em}p{3em} p{3em}p{3em} p{3em}p{3em} p{3em}p{3em} }      \\ \hline \hline
\multicolumn{14}{c}{ Molecules and terms } \\ \hline
ScH &  $^1 \Sigma$  & TiH &  $^4 \Phi  $  & TiO     &  $^3 \Delta$  & VH  & $^5 \Delta$  & CrN &  $^4 \Sigma$  & MnH &  $^7 \Sigma$  & FeH &  $^4 \Delta$  \\
ScN &  $^1 \Sigma$  & TiC &  $^3 \Sigma$  & TiF     &  $^4 \Phi  $  & VN  & $^3 \Delta$  & CrO &  $^5 \Phi  $  & MnO &  $^6 \Sigma$  & FeO &  $^5 \Delta$  \\
ScO &  $^2 \Sigma$  & TiN &  $^2 \Sigma$  & TiH$_4$ &  $^1  A_1  $  & VO  & $^4 \Sigma$  & CrF &  $^6 \Sigma$  & MnF &  $^7 \Sigma$  & FeF &  $^6 \Delta$  \\
ScF &  $^1 \Sigma$  & TiN &  $^2 \Delta$  & TiO$_2$ &  $^1 A_1   $  & VF  & $^5 \Pi   $  &     &               &     &               &     &               \\
    &               & TiN &  $^4 \Delta$  &         &               &     &              &     &               &     &               &     &               \\
    &               & TiN &  $^2 \Pi   $  &         &               &     &              &     &               &     &               &     &               \\ \hline \hline
\end{tabular} \  \
\caption{ \label{tab:2}
Test set of molecules, and molecular terms for each state.
Ground states are included for each molecule.
For TiN we include the ground state ($^2 \Sigma$) and three
excited states.
}
\end{table*}

To analyse the accuracy of the CEPPs in reproducing the properties of 
molecules, we select a test set of 24 small molecules in 27 states.
These molecules are composed of the transition metals Sc$-$Fe, together with the 
H, C, N, O, and F atoms whose CEPPs were previously 
reported.\cite{Trail_2013_pseudopotentials}
The test set was created by selecting the diatomic molecules for which the 
ground state term and dominant configuration is not in doubt, as described 
by Harrison,\cite{Harrison_2000} and keeping those for which single-reference 
CCSD(T) is stable (though not necessarily accurate).

More Ti molecules have been considered than for the other metals due 
to the technological importance of many of its compounds.\cite{Titanium_tech}
We focus on diatomic molecules, due to the computational cost of CCSD(T), 
but also include the TiH$_4$ and TiO$_2$ molecules to check for 
pseudopotential transferability beyond diatomics, and due to the importance 
of bulk polymorphs of TiO$_2$.\cite{Ma_2009}
Ground states are considered for each molecule, and three excited states 
are included for the TiN molecule to provide a further test of the 
transferability of the CEPPs.
While this test set is in no way complete, it covers a wide 
variety of chemical bonding behaviour.
The chosen molecules and terms are listed in Table~\ref{tab:2}.

\begin{figure*}[t]
\includegraphics[scale=1.00]{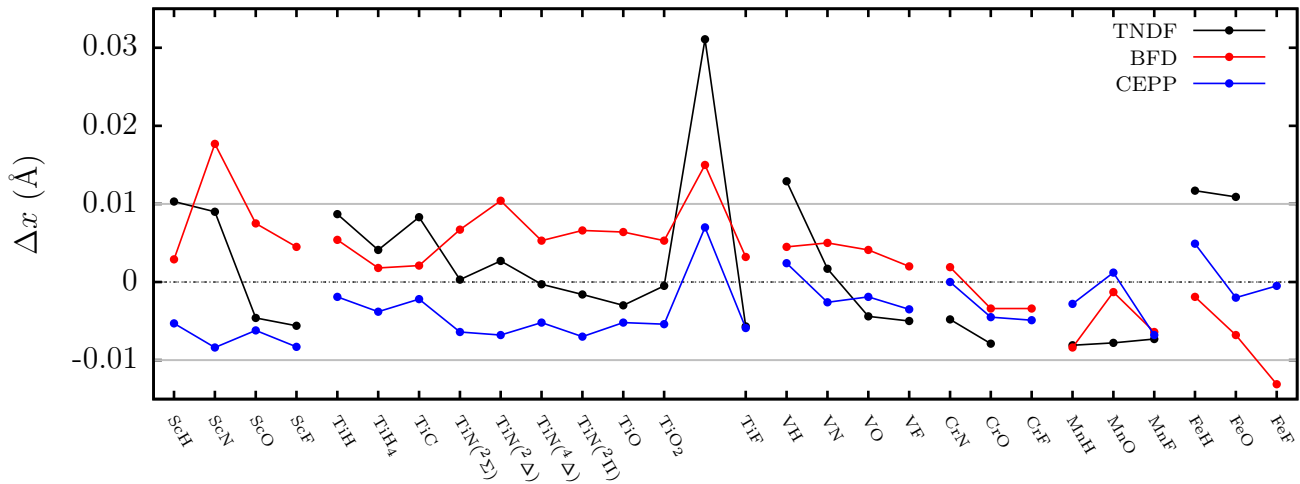}
\caption{ \label{fig3}
Deviation of spatial parameters from the baseline AE CCSD(T) results.
All parameters are bond lengths, except for TiO$_2$ for which a 
bond length and angle are present.
The latter is expressed as the arc-length of the bond angle 
on a circle of radius $1.0$\ \AA.
The AE and pseudopotential results are obtained by geometry optimisation as 
described in the text, using the TNDF and BFD pseudopotentials, and the CEPPs.
The horizontal grey lines indicate the boundaries for chemical accuracy.
}
\end{figure*}

We begin with geometry optimisation.
Overall, the CEPPs provide geometry parameters with a Mean Absolute Deviation 
(MAD) from the AE results that is 75\% of that for the BFD
pseudopotentials.
The maximum deviation of the CEPP geometry parameters from the AE results is 
47\% of that for the BFD pseudopotentials.
Perhaps the most important improvement is that for the CEPPs all
geometry parameters are within chemical accuracy (of $0.01$\ \AA) of
the relativistic AE results.
The small underestimate of the bond lengths (the mean deviation is 
$-0.003$\ \AA) can be ascribed to a combination of basis set error, 
the absence of a CPP, and transferability error.
Figure~\ref{fig3} shows optimal geometry parameters for all pseudopotentials 
as deviations from the baseline AE results.

\begin{figure*}[t]
\includegraphics[scale=1.00]{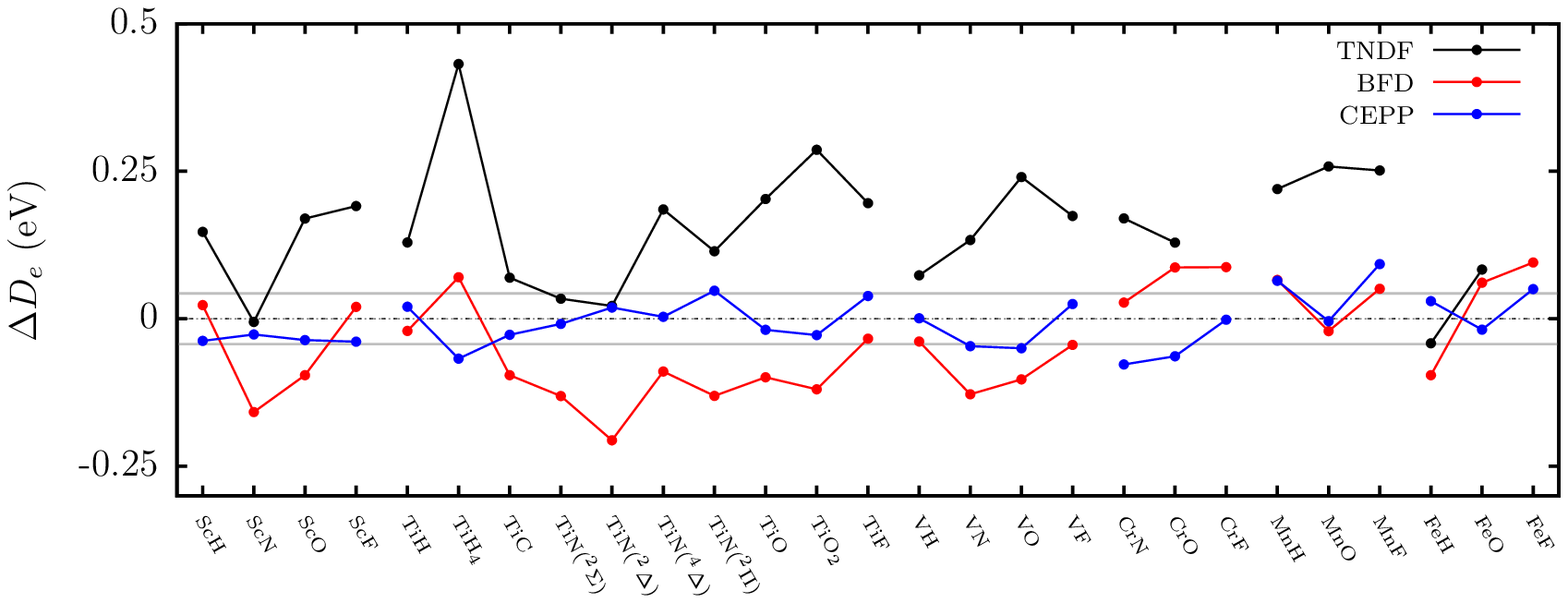}
\caption{ \label{fig4}
Deviation of well depths from the baseline AE CCSD(T) results.
The AE and pseudopotential results are obtained at the optimum geometries as 
described in the text, using the TNDF and BFD pseudopotentials, and the CEPPs.
The estimated error due to extrapolation to the complete basis set limit 
is less than $0.008$ eV for all molecules and pseudopotentials 
considered.
The horizontal grey lines indicate the boundaries for chemical accuracy.
}
\end{figure*}

The greatest improvement in accuracy provided by the CEPPs appears in 
the dissociation energies, $D_e$.
Overall, only 9 dissociation energies fall outside of chemical
accuracy (of $0.043$ eV), with a maximum deviation for MnN of only
$0.093$ eV.
The MAD is $0.035$ eV.
This MAD is 43\% and 22\% of the equivalent quantity for the BFD and TNDF 
pseudopotentials, respectively.
The improvement is just as pronounced when considering outliers, with a 
reduction in the magnitude of the maximum error to 45\% and 22\% of that 
for the BFD and TNDF pseudopotential results, respectively.
The MAD for the CEPPs is less than chemical accuracy, whereas for both BFD and 
TNDF, it is greater.
Similarly, 20 and 21 dissociation energies fall outside 
of chemical accuracy for BFD and TNDF pseudopotentials, respectively,
more than twice as many as for the CEPPs.
A further improvement is that the errors for the CEPPs shows less bias, with 
a mean difference between pseudopotential and AE results of $-0.006$ eV, 
compared with an average bias of $-0.038$ eV for the BFD pseudopotentials, 
and an average overestimate of $0.155$ eV for the TNDF pseudopotentials.
Note that the MAD error for the CEPPs is less than the average bias  
error for the BFD pseudopotentials.
Figure~\ref{fig4} shows dissociation energies, $D_e$, for all pseudopotentials 
as deviations from the baseline AE results.

We may quantify the importance of core-valence correlation by considering 
CEPP calculations with the CPP excluded, and by considering AE calculations
with no core excitations allowed (but with relaxed core orbitals). 
Excluding CPPs increases the MAD by only $0.003$ eV, and AE 
calculations without core excitations result in a MAD of $0.008$ eV. 
The small effect of excluding core-valence correlation strongly suggests 
that the improved accuracy of the CEPPs when compared with HF 
pseudopotentials is primarily due to better transferability, rather than 
the inclusion of core-valence correlation.

Harmonic zero point vibrational energies (ZPVEs) were obtained 
from numerical total energy 
derivatives at the optimised geometry.
Basis set convergence is easily achieved for the harmonic ZPVEs, 
hence we use the $n=T$ basis set for the CCSD(T) calculations, 
with the CPP included in the Hamiltonian.

For the calculated ZPVEs, the deviation of the pseudopotential results from AE 
results is well within chemical accuracy for all three pseudopotentials,
with a MAD from AE results of $16.9$, $9.0$, and $75.9$ cm$^{-1}$ for 
the CEPP, BFD, and TNDF pseudopotentials, respectively.

\begin{figure}[b]
\includegraphics[scale=1.00]{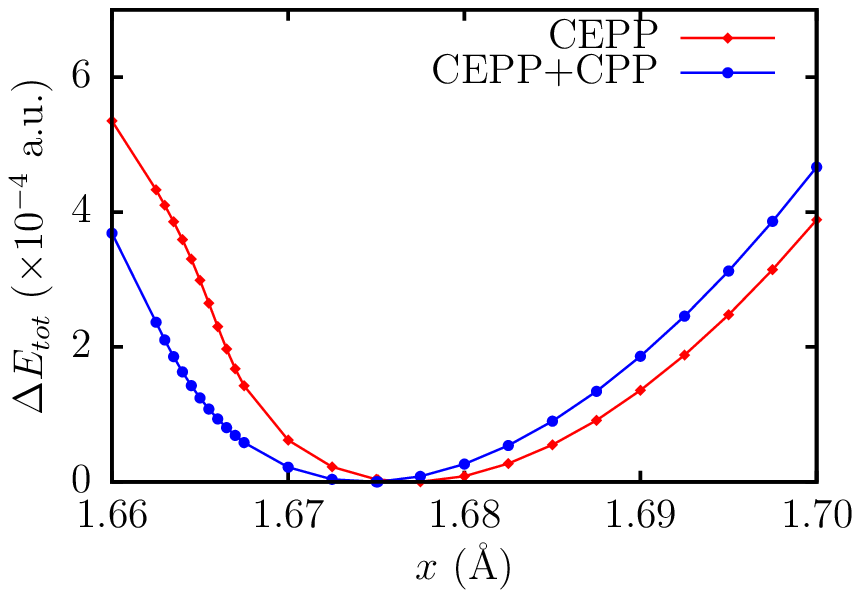}
\caption{ \label{fig6}
Total energy curves for TiC evaluated using the CEPP within CCSD(T), and the 
aug-cc-pVTZ basis set.
Results from calculations that both include and exclude the CPP are shown, 
with an offset added to the total energies to place the minimum 
at $\Delta E_{tot} = 0$.
}
\end{figure}

Closer examination of TiC demonstrates the errors that may arise if 
the CEPP is not used together with the CPP, primarily due to the relatively large 
core polarizability of both Ti and C.
Using the CEPP+CPP combination results in a ZPVE of $1199$ cm$^{-1}$,
in good agreement with the AE result of $1143$ cm$^{-1}$.
Excluding all of the CPP results in a ZPVE of $1350$ cm$^{-1}$, in poor 
agreement with the AE result.

Closer examination of the potential energy curves arising from the CEPP+CPP 
combination and the CEPP alone make this difference even more apparent.
Figure~\ref{fig6} shows such curves in the region of the equilibrium bond length, 
with an offset introduced to locate the minimum for both curves at zero energy.
The changes in equilibrium bond length and dissociation energy are 
insignificant compared with chemical accuracy, with the CPP 
decreasing the bond length by $0.002$\ \AA, and increasing the dissociation 
energy by $0.022$\ eV.
An anomalous variation of total energy with bond distance close to
equilibrium seems likely to be due to deficiencies in the
semi-empirical CPP description of the one-body core-valence
interaction.  
This semi-empirical component is not present when the CPP is included.

Basis set convergence issues are particularly apparent for TiC.
Experimental data are not available, and \emph{ab initio} values 
in the literature\cite{TiC_1996,TiC_2003} range from $805-889$ 
cm$^{-1}$.
Our AE, and pseudopotential, ZPVEs deviate from this range significantly.
This appears to be due to deficiencies in the basis sets used in previous 
calculations.
We can reproduce the AE CCSD(T) value of Hack \emph{et al.}\cite{TiC_1996} 
by employing the same contracted basis with only s, p, and d basis functions 
used by them, rather than the larger aug-cc-pVTZ-DK basis set that we use.
Neither relativistic effects, nor uncontracting the basis sets, reproduce 
this disagreement.
The experimental value of $1126$ cm$^{-1}$ quoted by Tomanari and 
Tanaka\cite{TiC_2003} also agrees well with our AE and CEPP result, but 
the relevance of this value is unclear.\cite{TiC_1974}

Overall, the results for the 24 molecules in 27 states provide a useful test of 
the performance of our CEPPs.
The choice of test set covers a modest, but important selection of atoms,  
and a variety of bonding types.
Given that previous assessments of the performance of the BFD and TNDF pseudopotentials 
have been somewhat limited in the number of molecules considered, our test set may be 
considered as relatively demanding.

The CEPPs do not introduce the non-linear core error of KS-DFT pseudopotentials, 
and include core-valence electron correlation, so we may ascribe all errors to 
the transfer of scattering properties over a large energy range, with no 
contribution from an inadequate description of correlation effects in the 
pseudopotential itself.
The results demonstrate the accuracy of CEPPs used together with an accurate 
description of electron correlation.

The choice of a small [Ne] core pseudopotentials is expected to be important for an accurate 
description of the 3d-transition metal atoms, especially to the left of the row.
Results for our test set obtained using the large ([Ar]) core TNDF 
pseudopotentials are not given here, but support this expectation.
They consistently show poor performance, with a MAD for geometry parameters, 
$D_e$, and ZPVE of $0.139$ \AA, $1.19$ eV, and $76$ cm$^{-1}$.
These values are an order of magnitude worse than those for the [Ne] core 
pseudopotentials, and far from chemical accuracy for both geometry parameters 
and dissociation energies.

\section{TiO$_2$ energies from plane-wave KS-DFT with CEPPs \label{bulkdft_tio2}}

\begin{figure*}[t]
\includegraphics[scale=1.00]{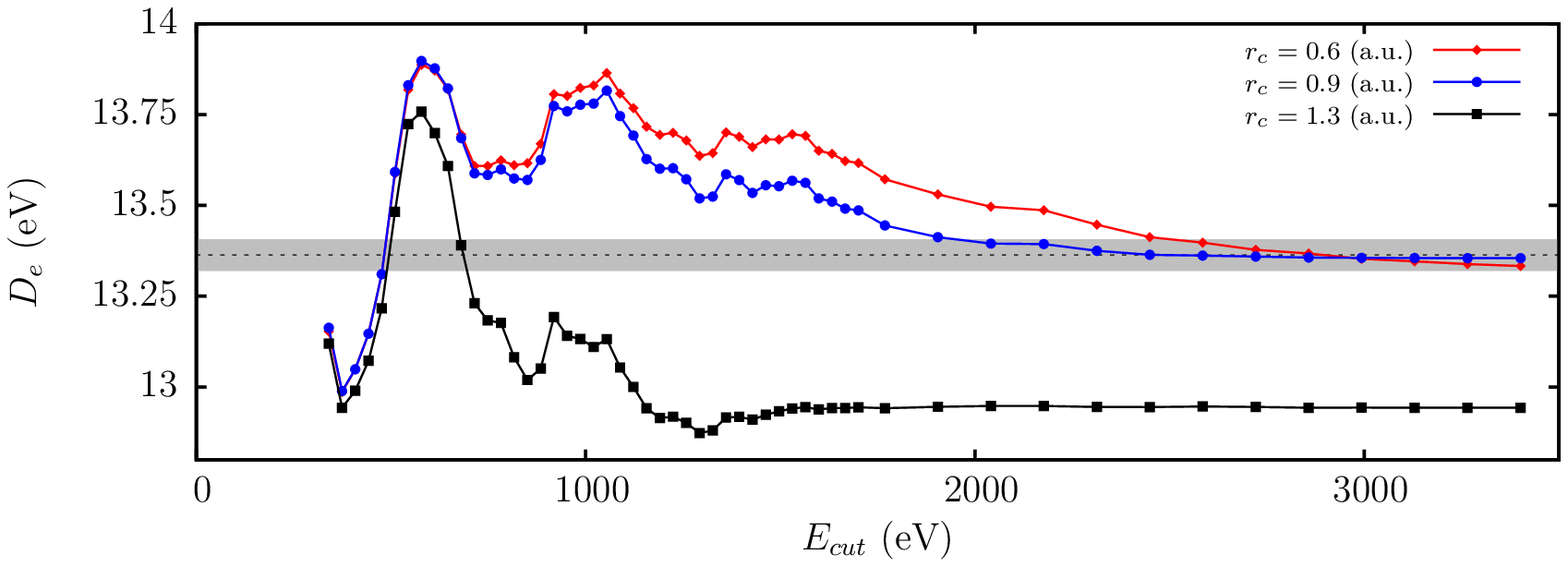}
\caption{ \label{fig7}
Convergence of the dissociation energy for a TiO$_2$ molecule with respect to 
the size of plane wave basis set, calculated using KS-DFT with
the PW91 exchange-correlation functional, and a KB representation of 
the CEPPs.
Results are shown for Ti CEPPs generated with different core radii 
for the d channel, at $r_c=0.6,0.9$, and $1.3$ (a.u.).
The Ti CEPP with $r_c=0.9$ (a.u.) shows the best convergence behaviour.
The grey region is within chemical accuracy of $D_e$ calculated using 
a Gaussian basis set, and relativistic AE CCSD(T).
}
\end{figure*}

A strength of QMC methods is their suitability for the description of 
correlated electrons in bulk systems.
Such calculations often involve periodic KS-DFT calculations with a 
plane wave basis, so it is desirable to validate 
the accuracy of CEPPs for this precursor to a QMC calculation.
We investigate the convergence properties of the dissociation energies 
that arise from calculations of this type.
Here we consider a TiO$_2$ molecule in a $(10$ a.u.$)^3$ cubic unit
cell.

The majority of plane wave KS-DFT implementations express the 
pseudopotentials in terms of projectors centred on each nuclear site.
This provides the pseudopotential operator in the Kleinman-Bylander 
(KB) representation\cite{Kleinman_1982},
\begin{equation}
\label{kb}
\hat{V} = V_{l_{max}} + \sum_l \sum_{ij} 
\left| \delta V_l \phi_{li} \right> B_{l,ij} \left< \phi_{lj} \delta V_l \right|,
\end{equation}
with $\delta V_l = V_l - V_{l_{max}}$.
The $\phi_{li}$ functions used as part of each projector are supplied 
for each channel of each pseudopotential, and the matrices for each 
channel are given by
$B_{l,ij}= \left< \phi_{li} | \delta V_l | \phi_{lj} \right>^{-1}$.
This separable form is less accurate than the semilocal spherical harmonic 
projector representation of Eq.\ (\ref{semiloc}), but the error is often small.
The primary advantage of the KB representation is a considerable reduction 
in computational cost.

For Ti and O we construct projectors using the numerical 
orbitals provided by atomic KS-DFT calculations with CEPPs.
For O we generate a single projector for each of the s and p channels 
from the $2s^22p^4$ ground state, with the d channel local.
For Ti we construct projectors using the 3s, 3p, 4s, and 3d orbitals 
present in the $3s^23p^63d^24s^2$ ground state.
A further projector is obtained using the 4p orbital from the 
$3s^23p^63d^24s^14p^1$ excited state, resulting in 5 projectors,  
with the f channel local.
For Ti, using two projectors for each of the s and p channels was 
judged to be desirable due to the significant contribution of the 
associated atomic orbitals to chemical bonding.

The CEPPs, and projectors, were used in the \textsc{CASTEP}\cite{castep} 
code (with a small modification to allow for more than one projector 
per pseudopotential channel) for unit cells containing TiO$_2$, Ti, and 
O.
Sums and differences of the resulting three total energies provide the 
dissociation energy for the molecule.

A PW91 exchange-correlation functional\cite{Perdew_1993} was used for 
both generating the projectors, and for the periodic calculations.
The energy cutoff for the plane wave basis set is denoted $E_{cut}$, 
and the geometry for the TiO$_2$ molecule was obtained by total energy 
minimisation with the largest $E_{cut}$ considered.

Figure~\ref{fig7} shows the convergence behaviour for three 
Ti CEPPs, each generated with a different value of $r_c$ chosen 
for the d channel.
All three result in well behaved convergence behaviour for $E_{cut} > 2200$ eV.
Results for $r_c=1.3$ a.u.\ show the fastest convergence with $E_{cut}$,
at the cost of poor transferability and a large offset error.
Results for $r_c=0.6$ a.u.\ show the slowest convergence with $E_{cut}$.
Results for the CEPP generated with $r_c=0.9$ a.u.\ provide the best 
compromise between convergence and transfer errors.

A region within chemical accuracy of the AE CCSD(T) value for $D_e$ 
is shown in Fig.~\ref{fig7}, which contains the KS-DFT results
for large $E_{cut}$ and small $r_c$.
While this supports our conclusion that the KB representation is successful 
for this CEPP, the very close agreement is probably fortuitous.
The variation of such a dissociation energy with different 
exchange-correlation functionals is expected to be significantly larger 
than chemical accuracy, and there is no reason to expect the PW91 functional 
to perform particularly well for this molecule.
We use $r_c=0.9$ a.u.\ in what follows, which is the value 
referred to in subsection\ \ref{implementation}.
Values of $r_c$ for the s, p and f channels were obtained similarly.

Incorrect `ghost' ground states may arise for the separable form
provided by the KB representation.\cite{Gonze_1991}
Transition metal pseudopotentials are particularly prone to this, 
due to their deep pseudopotentials, and hence we must check that the 
TiO$_2$ ground state is correct.  
We compare results that occur for a KB representation (Eq.~(\ref{kb})) 
with those for a semilocal representation (Eq.~(\ref{semiloc})).

For the KB representation, results were obtained as above, using the 
plane-wave basis with $E_{cut}=3400$ eV.
For the semilocal representation results were obtained using calculations 
equivalent to those in section\ \ref{results}, but with the CCSD(T) 
method replaced by KS-DFT.
The aug-cc-pV5Z basis set with no extrapolation was found to be 
sufficient for KS-DFT.
The PW91 exchange-correlation functional was used for both CEPP 
representations.

The total energies of the two CEPP representations agree well.
The KB and semilocal implementations provide total energies of 
$-90.859$\ a.u.\ and $-90.855$\ a.u.\, respectively, strongly 
suggesting that ghost states have not occurred.
Results for relaxed geometries show a similar close agreement, with 
the KB representation providing a bond length and angle of 
$1.632$\ \AA\ and $109.1^{\circ}$, 
and the semilocal representation providing values of
$1.641$\ \AA\ and $110.4^{\circ}$.
We conclude that ghost states do not occur for this system, and it 
is reasonable to expect that they will not occur for the rest of 
the CEPPs provided.

\vspace{0.5cm}
--------------------------------------------------------------------
\vspace{0.5cm}

\section{Conclusions \label{conclusions}}

We have extended a previously presented scheme for generating 
pseudopotentials to the 3d-transition metals, and including the 
relativistic effects expected to be important for these atoms.
These CEPPs are generated from explicitly correlated MCHF calculations, 
and core polarizabilities, and include intra-core and 
core-valence correlation effects.
We have generated such CEPPs for the atoms Sc$-$Fe.

The CCSD(T) method, together with these new CEPPs and some previously 
published CEPPs for light elements, have been applied to a range of 24 
mostly diatomic molecules in 27 states.

We have compared results obtained using the CEPPs with a baseline of
data for the same molecules and physical quantities, generated using
AE calculations.  This baseline AE data includes relativistic effects.
We also compared performance with the TNDF and BFD pseudopotentials
that are widely used in the QMC community.

The physical quantities compared are equilibrium geometries,
dissociation energies, and harmonic ZPVEs obtained.
The dissociation energies show a significant improvement in accuracy 
compared with the two HF pseudopotentials, with MAD and maximum deviations 
from the AE results for the 27 molecular states of: 
CEPP ($0.035$ and  $0.093$ eV),
BFD  ($0.081$ and $-0.206$ eV), and
TNDF ($0.158$ and  $0.432$ eV).
For the optimum geometries the CEPPs show a modest improvement over the 
BFD and TNDF pseudopotentials, but the improved accuracy is enough to 
ensure that all predicted geometries fall within chemical accuracy of 
the AE results for the CEPPs only.
For all molecules and pseudopotentials considered, the agreement of ZPVEs 
with AE values is well within chemical accuracy.

The CEPPs are generated from highly ionised atoms, hence the results 
obtained for the moderately large set of neutral molecules provide 
convincing evidence that such CEPPs are highly transferable.
The most demanding case is Fe.  We generate the Fe CEPP from a
Fe$^{+15}$ ion, and the resulting data for the FeH, FeO and FeF
molecules show an error that is not significantly greater than that
for the rest of the test set, or than that for the first row molecules
considered in a previous
publication.\cite{Trail_2013_pseudopotentials}

This very successful transfer of the CEPPs from ions to neutral systems suggests 
that the known poor transferability of norm-conserving KS-DFT pseudopotentials 
between different ionic states arises from the approximate inversion of the 
Kohn-Sham equations used to construct them.
Such an inversion process is necessarily approximate since it employs a 
linearisation of the exchange-correlation functional, and the exact 
functional is not available.

However, we do not conclude that our CEPP generation procedure will
necessarily be as successful for all atoms.
Although this may be the case, it must be tested atom by atom.
Furthermore, it may be argued that for the H$-$F and Sc$-$Fe molecules 
the core electrons replaced by the pseudopotential are more 
localised and tightly bound than for other atoms in the periodic table.
Whether this is significant remains an open question.
We conclude that the CEPPs generated here perform well, and
provide a consistent and significant improvement over the HF based 
pseudopotentials when used in correlated-electron calculations.

Tabulated and parameterised forms of the CEPPs described in this
paper, and orbitals for constructing KB projectors, are given 
in the supplementary material\cite{Supplemental}.

\begin{acknowledgments}
  The authors were supported by the Engineering and Physical Sciences
  Research Council (EPSRC) of the UK.
\end{acknowledgments}

%\appendix*

\end{document}